# ILC250 Cost Update - 2024

### Backup document for
#### "Status of the International Linear Collider"
#### (ESPPU2026 Input Document #275, arXiv:2505.11292 [hep-ex] )

G. Dugan, A. Lankford, B. List, S. Michizono, T. Nakada, M. Ross, H. Sakai,
S. Stapnes, N. Terunuma, N. Walker, and
A. Yamamoto (chair)

### ILC Cost-update Task Force

Submitted to ESPPU2026: 26 May 2025,
Submitted to arXiv: 31 May 2025

**Abstract:**

The International Linear Collider (ILC) was conceived as a global project for an energy-frontier electron-positron collider. It employs superconducting radiofrequency (SRF) and nano-beam technologies, with a center-of-mass energy rage of 250 GeV to be extendable to 1 TeV. Its cost was estimated for 500 GeV in 2013, based on the technical descriptions given in the 2013 Technical Design Report, for potential site in Americas, Europe, and Asia. After Japan's high-energy physics community proposed to host the ILC in Japan as a Higgs boson factory at 250 GeV in its first phase, a revised cost estimate for 250 GeV was conducted in 2017 for a Japanese site. However, due to global price increases and currency fluctuations that emerged afterward, the 2017 estimate is now outdated. A new cost evaluation has therefore been performed, accounting for global inflation trends, exchange rate shifts, and recent experiences in SRF based accelerators. The updated cost estimates are 6.78 billion ILCU (equivalent to 2024 USD) for the accelerator and conventional facilities, 196 billion JPY for civil engineering, and 10.12 thousand FTE-years for human resources provided by participating laboratories. These results are included in the ILC status report in May 2025, contributing to the ongoing 2026 update of the European Strategy for Particle Physics. This report provides the supplemental information.

Contact:     Akira Yamamoto (KEK)
e-mail:      akira.yamamoto@kek.jp

# 1. Introduction

The International Linear Collider (ILC) is an energy-frontier electron-positron collider based on superconducting radiofrequency (SRF) and nano-beam technologies . The ILC project originated with a global initiative, ILC Global Design Effort (GDE), to design a linear collider at a center-of-mass-energy (c.m.e.) of 500 GeV to be extendable to 1 TeV under the guidance of the International Committee for Future Colliders (ICFA) [1]. Subsequently, the Japanese high energy physics (HEP) community proposed to host the ILC in Japan as a global project, starting at a c.m.e. of 250 GeV as a Higgs factory [2]. The ILC construction cost for 500 GeV was estimated with the accelerator specification described in the Technical Design Report (TDR) published by the ILC-GDE in 2013 [3,4]. As a global project, the funding of the ILC was assumed to be mostly in-kind. Therefore, a special currency unit, ILC Unit (ILCU), was introduced to reflect this global character. It was defined as equivalent to the 2012 US Dollar, with the cost of items evaluated in various countries in their own currencies being converted with the Purchasing Power Parity (PPP) indices compiled by the Organization for Economic Co-operation and Development (OECD) [5]. With the Japanese HEP community's interest to start the ILC as a Higgs factory in its first phase, the cost for the ILC at c.m.e. 250 GeV was estimated in 2017 primarily based on the 2013 information. Table 1 summarizes the ILC250 baseline design parameters and Fig. 1 shows its general layout and technical configuration [6,7].  It is to be realized with power consumption of 111 MW corresponding to  2/3 or smaller than that of the LHC in operation.

Rapid global price increases and significant fluctuations in exchange rates have rendered, and  the 2017 cost estimate for the 250 GeV machine out of date. In view of the forthcoming European Strategy Particle Physics Update (ESPPU) in 2026, the International Development Team (IDT), which has been leading the ILC activities as a sub-panel of the ICFA, decided to update the 2017 ILC cost estimate [8]. For this purpose, the IDT set up a taskforce in spring 2024, and the taskforce has produced an updated cost estimate [9]. After the updated estimate was reviewed by an international review group [10], a concise summary of the updated cost estimate was included in an ILC Status Report submitted to the ESPPU 2026 process in March 2025 [11].

This document provides further details on the update of the cost estimate. After summarizing the methodology adopted from the original 2013 costing exercise, the work done for the 2024 update is outlined. This is followed by the updated cost estimate, including the additional studies for upgrading the energy to 500 GeV, for an additional beam delivery system with an interaction region, and for an electron driven positron source as a backup for the baseline undulator based positron source. Some of the comments brought up by the international review group are also addressed before concluding.

Table 1. ILC250 baseline parameters.

| Parameters | Value |
|---|---|
| $e^+ + e^-$ Beam Energy | 125 + 125 GeV |
| Luminosity | 1.35 x$10^{34}$ cm$^2$/s |
| Beam rep. rate | 5 Hz |
| Pulse duration | 0.73  ms |
| # bunch / pulse | 1312 |
| Beam Current | 5.8  mA |
| Beam size (y) at FF | 7.7 nm |
| SRF Field gradient | < 31.5 > MV/m (+/-20%) $Q_0 = 1\times10^{10}$ |
| Peak AC-plug power | 111 MW |
| Annual integrated power | 0.73 TWh / year |
| Length | 20.5 km |

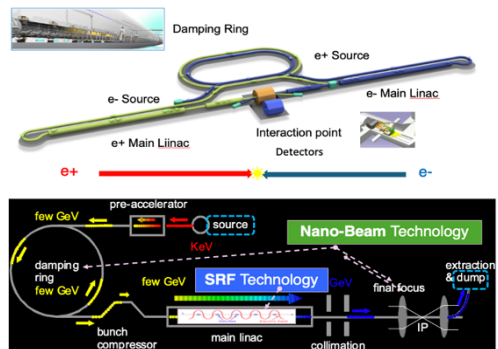

Fig. 1. ILC schematic layout and functional configuration.

# 2. Methodology of the ILC Cost Estimate

The ILC cost estimate for the TDR [3,4] is a full bottom-up cost estimate for operating at a 500 GeV centre-of-mass energy. It is based on around 2000-line items, going down to the level of the cost of individual components, such as cavities, couplers, or magnets, that are key cost drivers. The costs were evaluated and aggregated by an international team of experts and reviewed by an independent international expert panel in 2013.

In 2017, the cost estimate was updated for the staged construction of the ILC, starting with a 20.5km long accelerator operating as a Higgs factory at 250GeV [6,7]. This estimate was based on the TDR cost estimate, considering the effect of a reduced length and centre-of-mass energy as well as a number of minor design updates. The ILC250 cost estimate was given in 2012 prices, expressed in ILCU(2012), not adjusted for inflation, which was very low between 2012 and 2017.

## 2.1 Value and Labor

To account for differences in project cost accounting among countries, particularly in how labor costs were treated, the ILC cost model comprises two categories, following the example of other international projects such as ITER. One is "Value," which refers to the monetary value of goods and services procured from commercial vendors, expressed in ILCU. This includes accelerator components and system assembly acquired during the construction phase. The other category is "Labor," which denotes the total work performed by staff from participating institutions, including hired staff e.g. for installation work, measured in full-time equivalent person-years (FTE-years). In the ILC TDR cost estimate, Value and Labor are precisely defined as follows:

- **Value**, expressed in currency units, is defined for items procured from vendors: The Value of a component is defined as the lowest reasonable estimate of the procurement cost of an item with the required specification and in the necessary quantity, based on production costs in a major industrial nation.
- **Labor**, measured in person-hours or FTEs, is defined as "explicit" labor. This includes labor provided by the collaborating laboratories and institutions or purchased from industrial firms. It is important to distinguish this from a company's "implicit" labor embedded in the industrial production of components and contained (implicitly) within the purchase price. The implicit labor is included in the Value part of the estimate.

Value is expressed in ILC currency units, ILCU(2012), defined as the purchasing power of one U.S. dollar in the U.S. in January 2012 (see below).

## 2.2 Purchasing Power Parity (PPP)

PPP is an economic concept that relates prices for identical goods across countries. PPP currency rates permit the translation of local prices given in one currency into prices at a different location in the respective currency, serving as a regional deflator compiled through extensive surveys by the research arm of OECD and the European Union's statistical agency (Eurostat) and the World Bank [5,12].

## 2.3 Motivations for the Use of PPP Indices

For an international project primarily based on in-kind contributions such as the ILC, the use of PPP to compare prices and express them in a common currency unit is appropriate, because PPP reflects the cost for domestic production or procurement of the respective items. It is assumed that in-kind contributions would mostly be produced and purchased within the contributing regions. PPP conversion rates tend to vary more slowly than currency exchange rates, which are often driven by short term market fluctuations, particularly in times of economic or political upheavals such as the 2008 financial crisis or the 2020 Covid pandemic. The Value estimate in ILCU assigns the same value to in-kind contributions of identical items (e.g., cryo modules) regardless of the country of origin. This allows a fair comparison between vendors of similar capabilities in different regions, adjusted for local differences in purchasing power. The conversion of this Value estimate to a Cost estimate, where Cost is defined as the sum of actual expenditures by the participating in-kind contributors, is discussed below.

## 2.4 The scope of the cost estimate

The scope of the cost estimate encompassed the construction cost for the accelerator systems as detailed in the TDR, with a centre-of-mass energy of 500 GeV (or 250 GeV for the ILC250). Detectors were excluded from this cost estimate and a separate cost estimate for the two detector concepts SiD and ILD is given in [13]. However, the following items were included:

- detector assembly buildings,
- underground experimental halls, and
- detector access shafts.

Computing equipment for accelerator operation was included, while computing installations for detector operation, data taking and analysis were excluded. The cost estimate covers the 9 year construction time until the start of commissioning. Therefore it excludes the following items;

- costs for project engineering and design,
- R&D prior to construction authorization, and
- commissioning, pre-operation, operation, and decommissioning.

Also not included are

- taxes,
- contingency,
- escalation during project construction,
- land acquisition, and
- site activation (external roads, water supplies, power supply lines).

Costs for upgrading the machine beyond 500GeV are not included, except the cost of those systems that would be very difficult to provide after construction of the 500 GeV machine, in particular the main beam dumps and the beam delivery system (BDS).

## 2.5 Data for the TDR cost estimate

The TDR cost estimate was based on approximately 2000 individual line items. For each item, the following details were recorded;

- the number of items,
- unit costs,
- basis of estimate,
- cost category,
- cost uncertainty,
- country of origin, and
- local currency.

To protect future tendering and bidding processes, detailed information on individual items was kept confidential, and only sum values were made public. To give insight into the cost structure and cost drivers, breakdowns according to accelerator areas or technical systems were provided.

## 2.6 Value matrix

A two-dimensional value matrix was an important tool. In the value matrix, various technical components were allocated according to "Technical system" (SRF-related, magnets, vacuum and others), and "Accelerator-area system" (such as Sources, DR, RTML, ML, BDS and others). For example, the SRF-related technical systems were distributed in Sources, DR, RTML, and ML, and magnets were distributed in every accelerator area system. These matrices enabled costs to be presented from two perspectives: by technical system and by accelerator area system.

## 2.7 Cost premium

The cost uncertainty, or cost premium, is defined as the procurement risk of the Value estimate. It is given by the difference between the cost estimate at 50% confidence level (median) and that at 84% confidence level, corresponding to a 1σ uncertainty for normally distributed data [3,4]. The uncertainty for each cost element depends on the nature, quality and maturity of the element. In the TDR estimate, uncertainties were evaluated by cost estimators for each cost element, based on defined guidelines, considerting:

- maturity of the item's design (conceptual, preliminary, or detailed),
- level of technical risk involved in the design and manufacture of the item,
- impact of delays in this item on the project schedule (critical-path impact, non-critical-path impact, no schedule impact on any other item),
- source of the cost information (engineering estimate based on minimal experience, engineering estimate based on extensive experience, vendor quote, industrial study, catalogue price), and
- the extent, if any, of cost scaling to large quantities.

The cost premiums of line items were added linearly, assuming full correlation between uncertainties, a conservative approach. The resulting relative uncertainties varied for different technical systems, and the overall uncertainty was evaluated as 25 % in the TDR.

## 2.8 The conversion of the Value estimate

The process for converting the Value estimate to a national cost estimate was documented in [14]. Here, "cost" refers to the total amount of monetary expenditure required to procure all items contained in the Value estimate. Items such as cryomodules that could be procured from multiple vendors in the regions participating in the project (assumed to be North America, Europe and Asia, mainly Japan for the ILC) could vary in price due to differences in regional buying powers. Depending on the procurement policy of in-kind contributors, items could be acquired locally, possibly at a higher cost, or on the world market. Because procurement is decentralized in an in-kind project, actual costs were not governed by a central policy. Therefore, care had to be taken in converting the Value estimate provided in ILCU into any currency, as any such conversion implicitly relied on a procurement model.

## 3.   Procedure for the ILC250 Cost Update - 2024

Between 2017 and 2024, the worldwide economic situation and price levels significantly changed due to the Covid pandemic and the war in Ukraine. In addition, planning of various civil construction projects in Japan progressed, and procurement results from several accelerator projects such as Eu-XFEL, LCLS-II and others based on TESLA technology became available [15,16]. These developments motivated an update of the cost estimate for the ILC250 in current prices.

In order to preserve consistency with the 2017 cost estimate and because the funding model for the ILC as a project based on international in-kind contribution project remains unchanged, it was decided to retain the scope and methodology of the original cost estimate used at the TDR, in particular the use of PPP to convert costs from local currencies into Value expressed in a single currency unit ILCU.

In the ILC250 cost evaluation, the cost categorized as 'Value' includes the cost for the Accelerator systems (ACC), which consist of superconducting RF related (SRF) and other conventional accelerator technologies, as well as Conventional Facilities (CF) for accelerator utility services, and Civil Engineering (CE) for underground and surface construction. The Value was divided such that ACC and CF were to be globally shared, while CE was to be locally contributed by Japan as the expected host state. The scope of the cost estimate update encompassed the construction cost for the accelerator systems as defined in Section 2.

In the 2017 cost estimate update, the reduction in the number of main linac components—resulting from the decrease in energy from 500 to 250 GeV—had already been implemented. For the 2024 update presented here, extensive efforts were made to develop new cost estimates for the SRF, CF, and CE technical systems that accounted for over 75% of the ILC250 total cost in 2017. These estimates were developed in communication with worldwide industrial partners and collaborating laboratories, including visits and meetings in person to directly receive their up-to-date cost estimates, as of 2024. Therefore, the effect of inflation between the 2012 TDR estimates and this 2024 ILC250 estimate update is already reflected for those items. The remaining fraction accounting for less than 25% of the ILC250 cost in 2017, was updated by scaling by the reductions in the numbers of components, by incorporating the effect of worldwide inflation, and by applying the same cost estimation methodology as applied in the TDR cost estimate.

For the 2024 update, the Value estimate was given in 2024 prices and given in ILC currency units ILCU(2024) defined as the purchasing power of 1 USD in the United States, as of January 2024. PPP currency conversion rates were based on the published OECD values for 2023 with a small extrapolation to Jan 2024 derived from the relative inflation indices [5,12,14,17,18]. For Germany, Japan, and Switzerland, data published by OECD were used [5,17]. For China, for which there is no OECD data, data published by the World Bank were applied [12]. The PPP and exchange-rate (Ex) indices used in 2024, compared with those in 2012, are summarized in Table 2. The PPP indices for Machinery and Equipment (M&E) were used for the accelerator technology (ACC) and conventional facilities (CF). In accordance with the TDR procedure, the prices of superconducting material used in the SRF cavity were converted to USD with the currency market exchange rate, because raw material commodities generally have world market prices in USD. The PPP indices for Civil Engineering (CE) construction are listed as information only, because that the CE estimate is specific to an ILC in Japan.

Table 2. PPP and exchange-rate (Ex) indices used in the cost in 2024.

| | US ILCU = USD | Germany ILCU -> EUR | Japan ILCU -> JPY | Swiss ILCU -> CHF | China ILCU -> CNY |
|---|---|---|---|---|---|
| | 2012 / 2024 | 2012 / 2024 | 2012 / 2024 | 2012 /2024 | 2012 / 2024 |
| **M&E** | 1 / 1 | 0.923 / 0.91 | 127.3 / 129.6 | 1.480 / 1.05 | -- / 8.83 |
| **SC Material.** | 1 / 1 | n/a | 76.9 / 149.1 | n/a | -- / 7.10 |
| **(CE-cost.)** | (1 / 1) | (0776 / 0.77) | (109.3 / 99.5) | (1.303 / 0.73) | (n/a) |

Variations in the inflation indices for M&E in each region for the US, Japan, Germany, and Switzerland from 2012 through 2024 are shown in Fig. 2, normalized to 100 % in 2012 [18]. Significant inflation was experienced worldwide between 2017 and 2024, and it strongly contributed to the cost increase in every technical system.

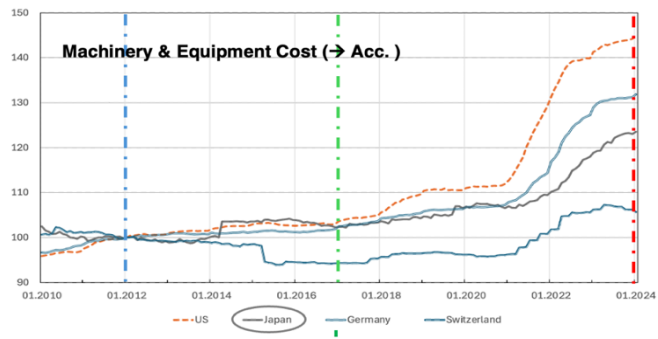

Fig. 2. Time evolution of inflation indices for machinery & equipment, normalized to 100% in Jan. 2012.

The ILC250 (2024) cost update proceeded as follows:
1) Prepare an initial ILC250 cost Value matrix:
   - by using the cost methodology associated with known changes between 2012 and 2024, such as production scaling, inflation, changes in purchasing power (in terms of PPP)
2) Collect new cost inputs as of 2024 for :
   - SRF cost, with inputs from industry and laboratories worldwide, and averaged to be properly conservative, covering the overall cost fraction of 80% to the total SRF cost in ILC250 (2017),

- CE and CF costs, with inputs from consulting companies and laboratories in Japan, following Japanese strict guideline, covering 100% of the CE and CF cost in ILC250 (2017),
3) Update the Value matrix with the new input numbers,
  - The total new cost inputs replacing >70% of the overall ILC250 (2024) cost,
4) Finalize the new ILC250 Value Matrix as of 2024.

Figure 3 summarizes workflows of the ILC cost estimate update in three steps: (i) ILC500 (TDR) published in 2013, (ii) ILC250 proposed in 2017, and (iii) ILC250 cost updated in 2024, where the CE and CF cost estimate study was dedicated to Japan, while the accelerator technical system study was kept worldwide.

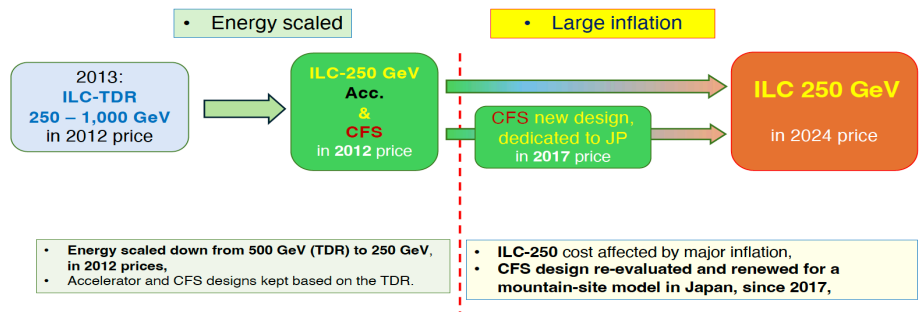

Fig. 3. Workflow of the ILC cost estimate starting from ILC-TDR, to ILC250 in 2017, and finally to ILC250 in 2024.

## 3.1 SRF-related Accelerator technical systems

The main cost drivers of the accelerator systems are SRF systems for the Main Linac, in particular the SRF cavities and Niobium raw material, cryomodules, high level RF (HLRF) equipment such as klystrons, and the cryogenic plants. The 2024 Value estimate updated for these items was based on several studies, with new cost estimates from studies by industrial partners and actual procurement experience from the Eu-XFEL and LCLS-II projects. In the cost estimate update process in 2024, new SRF-related cost estimate inputs were collected by the ILC cost update task force in close communication with well experienced industrial partners and expert laboratories and institutes, covering the sub-categories of superconducting material, SRF cavity and cryomodule production, HLRF and cryogenics, including a number of visits to vendor factories worldwide with deep discussion in person to directly receive the most updated information. These efforts are summarized in Table 3.

Table 3. ILC SRF-related cost inquiry and quotations collected in 2024 prices.

| Category | Productions | Technical Notes | # units: ~1/3 shared, industrial production for ILC250 | New inputs in 2024, collected from vendors / projects, laboratories in global, 3 regions: Americas [AM], Europe [EU], and Asia [AS] | |
|---|---|---|---|---|---|
| SC material | Nb300-sheets | 0.265 (sq)x0.0028m | 60,000 (~300 tons in total) | AM, AS / Eu-XFEL, LCLS-II | |
| Cavity | 1.3 GHz resonator | 9-cell (~1m), E: 35 MV/m (+/-20%) | 3,000 | EU, AS / Eu-XFEL, LCLS-II | |
| | Input power coupler | 1.3 GHz, 1.65 ms, 5 Hz, 600 kW | 3,000 | EU, AS / Eu-XFEL, LCLS-II | |
| | Tuner | Slow tuner tuning range: >600 kHz Fast tuner, tuning range: >1kHz | 3,000 | AS / Eu-XFEL, FNAL, KEK | |
| | Magnetic shield | Permalloy sheet, 1-layer outer shield | 3,000 | AS / Eu-XFEL, KEK | |
| SC mag. | SC-mag + BPM | 40T/m, 0.9 m (ap), 0.25/1m (length) | 110 | AS / LCLS-II, KEK | |
| Cryomodule- | Thermal ins. Vac. V. | 1 m (dia.) x 12.5 m (length) | 330 | EU, AS / Eu-XFEL, LCLS-II | |
| | Assembly with cavity | Assembly work hosted at laboratories | 330 | – / Eu-XFEL, CEA | |
| HLRF | Modulator | 10 MW, 120kV, 140A, 1.65ms, 5Hz, | 80 | (TDR data scaled) | |
| | Klystron | 1.3 GHz, 10 MW, 1-klystron / 26 cav. 5.8 mA, 32.7 MW, 1.65 ms, 5 Hz | 80 | AS / KEK | |
| | Power Distr. System | Distr. w/ <8% average lose in PDS | 80 | (TDR data scaled) | |
| Cryogenics | He Cooling System | ~ 20 kW @ 4.5 K, ~ 6 plants, and ~ ≤ 2 kW @ 4.5 K , 2–4 plants | ≤ 3 ≤ 2 | – / LCLS-II, LHC, CERN | |

For these updated cost estimates, the line items in the original TDR cost estimate were identified and the Value estimate for these items was replaced by the new input numbers. The relative cost uncertainty associated with these items was retained, as the TDR estimates were typically of the same quality (industrial studies by qualified vendors, corroborated by Eu-XFEL procurement experience).

The number of items to be installed was retained from the 2017 cost estimate. An 11% overproduction of cavities over the number of installed cavities, which corresponds to a 90% production yield, was considered, as was the number of additional tests following from an 80% assumed first pass yield and 25% retreatment rate. The number of cryomodules included in the costing covers the 6% additional cryomodules in the main linac. For large scale production of items such as cavities and cryo modules, a 95% learning curve was used, if necessary, to scale procurement results to the number of items required for the ILC250; industrial studies were asked to provide quotes for the corresponding quantities, and were scaled to a three vendor model in the ILC250 cost estimate in 2024, as properly conservative assumption for the ILC program which will be globally contributed from three major regions.

The Value estimate of items for which no updated estimates were obtained was corrected for inflation by first converting the 2012 price in ILCU(2012) back to the original cost estimate in the corresponding local currency (USD, EUR, JPY, or CHF). Then the appropriate inflation factor [18] was applied, and the result was converted to ILCU2024 with the 2024 PPP conversion rate given in Tab. 1. For the cost uncertainty, the original TDR uncertainty was added in quadrature with an estimated uncertainty for the inflation correction from 2012 to 2024. The estimated uncertainty was determined by an analysis of the spread of inflation data between the subcategories used in the determination of the machinery and equipment inflation index. The inflation factors range between 1.057±0.145 for Switzerland and 1.453±0.191 for the U.S. After the quadratic addition, the impact of the uncertainty from inflation is small compared to the typical uncertainties of most line items of 25% to 35%, depending based on the estimate. This procedure was applied to the estimates for conventional accelerator systems such as magnets and power supplies, vacuum system, diagnostics and control system, dumps, and smaller, area-specific components for the sources or undulators and wigglers in the damping rings.

## 3.2 Conventional Facility

For the CF estimate, the design and cost update followed the study conducted in cooperation with industrial consultant partners and experts of experienced laboratories and universities. It should be noted that significant inflation for raw material cost such as Cu conductor material in the past several years, as shown in Fig. 4, significantly affected the cost increase on top of general inflation effects [19]

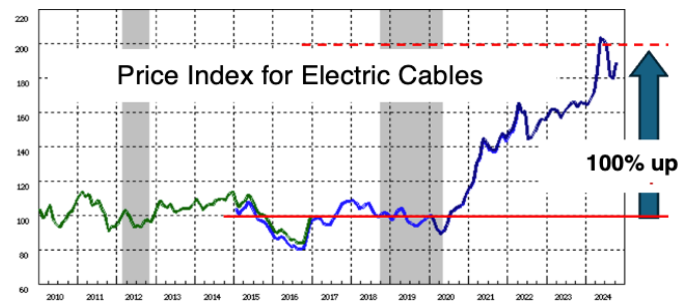

Fig. 4. Inflation for Cu cable price in 2020-2024.

## 3.3 Civil Engineering

For the CE estimate, the design and cost update followed the Japanese Government guidelines and the national tunnel costing standards. These standards, updated annually by the Ministry of Land, Infrastructure and Transport [20], provide a strict, reliable and well-established framework for tunnel construction in Japan. Once the design is fixed, the cost is precisely determined by a dedicated method, provided the site is well scoped.

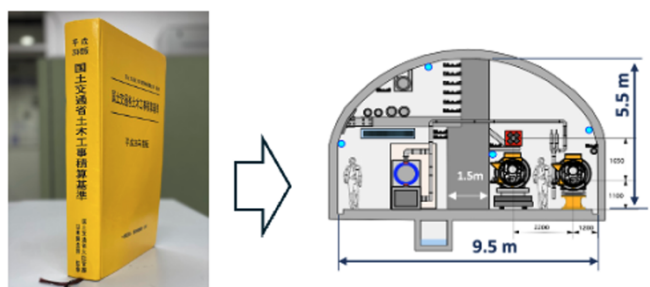

Fig. 5. CE work, following JP government guideline.

The plan was reviewed by the Japan Society of Civil Engineers (JSCE), concluding the plan to be technically appropriate [21]. Since the CE construction is assumed to be fully contributed by Japan, the updated CE cost is kept in Japanese Yen to be free from unnecessary currency exchange rate variation, year by year. This was also recommended by the International Cost Review Committee [10]. In contrast, the cost estimates for the ACC and CF subsystems are given in the ILCU. The CE cost has been also generally affected by inflation particularly in past 10 years, as shown in Fig. 6 [22].

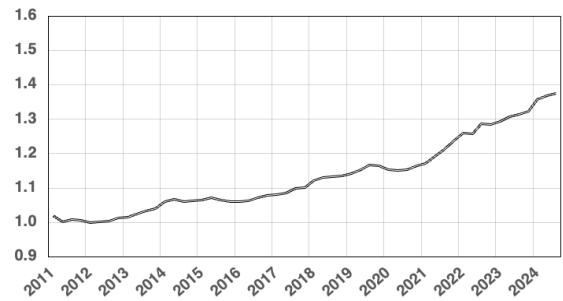

Fig. 6. CE inflation index increases since 2012.

## 3.4 Additional Studies
Additional cost estimate studies have been made on extra costs required for
1) energy upgrade from 250 GeV to 500 GeV,
2) an additional beam interaction point (for a total of 2 IPs), and
3) the electron-driven positron source option as a backup.

# 4. ILC Cost Update Results - 2024
## 4.1 ILC250 Cost Update
The ILC250 cost update in 2024 and its breakdown are summarized in Tables 4a and 4b together with corresponding pie charts. The high-level cost updates are summarized in Table 5 in comparison with the estimates for ILC500 in 2012 and ILC250 in 2017. The accelerator construction costs are categorised as the sum of costs for accelerator components (ACC: 5.40 billion ILCU) and conventional facilities (CF: 1.38 billion ILCU) resulting in a total of 6.78 billion ILCU. The cost for civil engineering (CE) is 196 billion JPY.

The combined ACC+CF cost has increased by approximately 60% since the previous estimate (2017). Of this increase, about 35% is due to worldwide inflation and the remainder reflects the change in production scaling, averaging multiple vendor cost-estimates, exchange rate variation, design updates and other smaller changes. The CE cost has risen by approximately 50%, where about 30% is due to cost increases in Japan and the remainder is due to recent design updates and enhancements.

The cost-estimate uncertainty linked to procurement (so-called cost premium) has been evaluated to ensure an 84% confidence level in covering the total project cost, corresponding to a one sigma cost. The uncertainties vary for different technical systems, and the overall uncertainty is evaluated as 29% of the total estimated cost, including uncertainty due to inflation, in the ILC250 (2024) cost estimation.

The estimate for human resources (Labor) includes both laboratory staff and installation workers, potentially under contract. This is expressed in FTE-years (full time equivalent person -years equivalent to). estimates remain unchanged from the 2017 evaluation.

As recommended by the experts panel on project implementation, a contingency budget to address unforeseen circumstances needs to be established centrally at the start of the project [23]. This contingency is estimated at a level of 10% of the total project cost. It is important to note that this represents a limited contingency, like to those typically used in European infrastructure projects, rather than a full contingency allowance.

**Table 4, ILC250 Cost (Value) Breakdown updated in 2024**

**4a) Technical-system oriented:**

| Tech. Systems: | Sub-Technical System | B.D. [B ILCU] | Value [B ILCU] | B.D. [B JPY] | Value [B JPY] |
|---|---|---|---|---|---|
| **SRF - related** | | | **3.69** | | |
| | 1.3-GHz Cavity (SC-mat, Fab., Surface) | 1.689 | | | |
| | Cryomodule (Parts, Assembly) | 0.701 | | | |
| | L-band HLRF (Modulator, Klystron, PDS) | 0.635 | | | |
| | Cryogenics | 0.661 | | | |
| **Conv. Acc. Tech.** | | | **1.71** | | |
| | Magnet & Magnet Power-Supply | 0.642 | | | |
| | Vacuum | 0.145 | | | |
| | Beam-dump & Collimator | 0.071 | | | |
| | Instrumentation | 0.183 | | | |
| | LLRF, Control, & Computing-Infra. | 0.483 | | | |
| | Others (Area-specific., Installation-Equip.) | 0.188 | | | |
| **Conv. Facility (CF)** | | | **1.38** | | |
| | Electrictrical Distribution | 0.835 | | ) | |
| | Cooling & Ventilation | 0.368 | | | |
| | Others (alignment, safety, etc.) | 0.179 | | | |
| **Civil Engineering (CE)** | | | | | **196** |
| | Underground (Tunnel, Cavern, Access T.) | | | 158 | |
| | Surface Build. (IP, Acc.-Site, Main Campus) | | | 38 | |
| **Sum (Acc+CF)** | | | **6.78** | | |
| **Sum (CE)** | | | | | **196** |

**4b) Accelerator-Area oriented:**

| Accelerator Areas: | Note | | Value [B ILCU] | | Value [B JPY] |
|---|---|---|---|---|---|
| **e- source** | | | 0.20 | | |
| **e+ source** | Undulator-driven e+ source | | 0.34 | | |
| **DR** | Damping Ring (two rings) | | 0.47 | | |
| **RTML** | Ring-to-MainLinac | | 0.69 | | |
| **ML** | Main Linac | | 3.14 | | |
| **BDS & IR** | Beam Delivery System & Int. Region | | 0.24 | | |
| | | | | | |
| **CF** | Conv. Facility (cooling & ventil.) | | 1.38 | | |
| **CE** | Civil Eng. (under-gr., surface) | | | | 196 |
| **Sum (Acc+CF)** | | | **6.78** | | |
| **Sum (CE)** | | | | | **196** |

**Note 1:**
- The cost-update for ILC250 baseline design determined by Luminosity : $1.35 \times 10^{34} \text{cm}^{-2}\text{s}^{-1}$.
- 1 ILCU = 1 USD (Jan. 2012) for ILC500 (2012) and 1 ILCU = 1 USD (Jan. 2024) for ILC250 (2024).

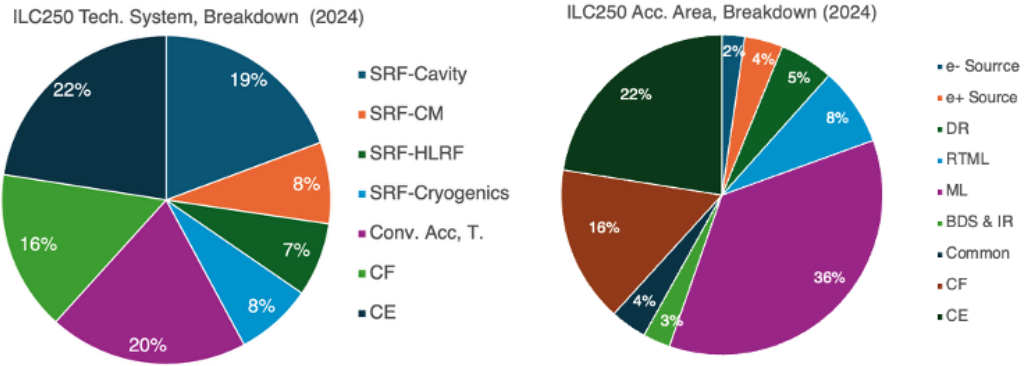

| Note 2: |
| --- |
| • **Fractions of the ILC 250 (2024) costs, Technical and Acc. Area Breakdown.**<br>➢ All Cost ratio plotted in ILCU, (i.e. CE-cost converted to ILCU with 2024-CE-constr. Index).<br>➢ **More than 70%** of the cost updated **with new inputs in 2024,** (i.e. > 80% of SRF, and 100 % for CF and CE). |

**Table 5. ILC Cost Update comparisons: ILC500 (2012), ILC250 (2017), and ILC250 (2024).**

| Cost Estimate/Updates (Year) | ILC500 (2012) | | ILC250 (2017) | | ILC250 (2024) | |
| --- | --- | --- | --- | --- | --- | --- |
| **Accelerator Construction (in 9 years)**<br>**Value: Acc. + Conv. Facility (CF : global)**<br>**Civil Engineering (CE : JP specific)** | **[B_ILCU]**<br>6.52 | **[B JPY]**<br><br>160 | **[B ILCU]**<br>4.24 | **[B JPY]**<br><br>129 | **[B ILCU]**<br>6.78 | **[B JPY]**<br><br>196 |
| **Breakdown:**<br>   **Acc-SRF related**<br>   **Conv. Acc. Tech. (mag., vac, and others)**<br>   **CF (utility service): Electric., cooling, ventil.** | 4.32<br>1.39<br>0.91 | | 2.34<br>1.20<br>0.71 | | 3.69<br>1.71<br>1.38 | |
| **Labor (HR): Laboratory staff**<br>   **Installation worker** | 10.12 [k FTE-yrs]<br>3.35 [k FTE-yrs] | | 7.47<br>2.65 | | 7.47<br>2.65 | |
| **Acc. Operation (/year)**<br>   **Value (Electricity, Cooling, etc.)**<br>   **Labor (HR):** | 0.39 [BILCU/yr]<br>850 [FTE] | | 0.32<br>638 | | 0.41<br>638 | |
| **Uncertainty (cost premium) [3,4]** | 25% | | 25 | | 29 | |
| **Contingency (common fund reserve) [21]** | 10% | | 10 | | 10 | |
| **References** | [3,4] | | [6,7] | | [9,11] | |

## 4.2 Additional Studies

Rough cost estimates are made for the following three cases:

- The cost to upgrade the ILC energy from 250 to 500 GeV is estimated as 3.9 to 4.2 billion ILCU for ACC+CF plus 55 billion JPY for CE.
- The cost to add a second beam-interaction point is estimated to be of the order of 0.5 billion ILCU. This estimate includes a second beam delivery system (BDS) and detector hall, but not the CE work nor the beam splitting systems needed at both BDS upstream ends. Dedicated design work would be needed to improve the accuracy of this estimate.
- The differential cost for the electron-driven positron source to the cost of the undulator-based positron source (given in Table 4b) is estimated to be 0.20 billion ILCU for ACC+CF, plus 12.5 billion JPY for CE to provide a second dedicated tunnel, which must be ready before starting beam operation.  This solution might be needed as a backup plan, in case the undulator-driven positron source technology requires more time to mature.

# 5. Responses to the comments from the review group

During the international cost review, particular attention was given to three key points: acceleration gradient margin; validity of the estimate of civil engineering cost, and choice of baseline positron source. The following subsections address each of these points.

## 5.1 Acceleration gradient margin

The international review group was concerned by the technical assumption of an average acceleration gradient of 31.5 MeV/m for the SRF cavities, because this performance has not yet been demonstrated in an operational machine, and because of the risk that the design energy will not be achieved.

The ILC TDR design includes several measures to address energy margin. First, it specifies a design gradient of 35 MV/m ±20% for bare SRF cavities. This performance target was achieved with a yield of 94% in the 9-cell bare cavity tests, as reported in the TDR. The operational gradient of the SRF cavities after assembly into the ILC main linac cryomodules was set to an average of 31.5 MV/m, incorporating a 10% reduction from the bare-cavity gradient to ensure operational margin. Second, as a feature of the ILC main linac SRF system design, the operating gradient of cavities can be individually adjusted by using local power distribution systems that provide the flexibility to remotely and independently optimize the power delivered to each individual cavity to account for the expected ± 20% spread in gradient performance [3]. The ILC250 design follows this specification. Third, the specified number of installed cryomodules provides a 6% energy margin. Fourth, an 11% production margin is included in the costing to account for an expected 90% cavity production yield resulting from the higher gradient requirement compared to the Eu-XFEL specifications. The concern of the review group is well appreciated. If needed, the existing energy margin can be increased in a number of ways. The energy margin can be increased by increasing the number of cavities and cryomodules installed in the main linac. Our study shows that 21 additional cryomodules (7x3 CM → 7x (9+8+9) = 182 cavities) per linac can be accommodated in the existing 280-meter-long drift space in each of the e+ and e- main linacs, as illustrated in Figs. 7a and 7b [6]. This space is part of the drift region used to adjust electron-positron collision timing for the undulator-driven positron source. These additional modules provide approximately a 5% increase of the energy margin, supplementing the existing 6% margin in the current ILC250 design, for a total energy margin exceeding 10%. In addition, the 11% production margin can be increased if the yield losses are less than the budgeted 90%. The yield will be established by a pilot production run of a sufficient number of cavities and cryomodules during the project preparatory phase, in advance of construction start.

Increasing the number of surface preparations and treatments in the cavity fabrication process to improve the production yield in a cost-effective way can also be considered. It is well known that the surface treatment accounts for about one third of the total cavity production cost including the superconducting material. This possibility will be further studied.

Preliminary evaluation shows that the additional cost required for both installing additional cryomodules and increasing the number of surface treatments will remain within 5% of the total cost of the ILC250. Further details will be studied during the coming years with accelerator technology R&D followed by the preparatory phase for the ILC Engineering Design Report (EDR) [24-26].

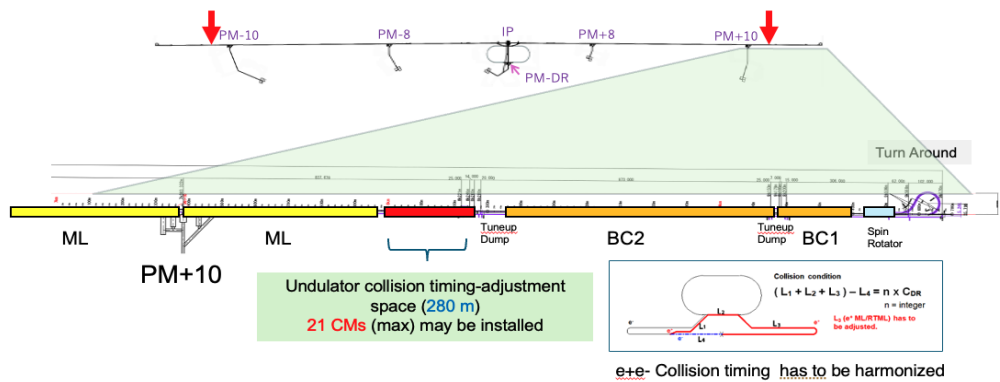

Fig. 7a. Available tunnel space remaining for additional CM installation.

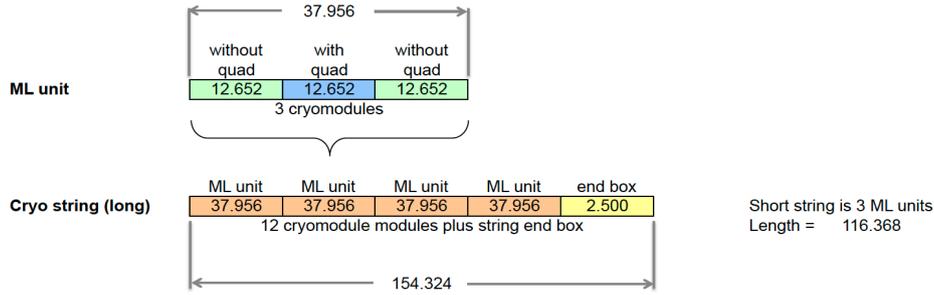

Fig. 7b. ILC ML CM string configuration; ML-unit consisting of 3 CMs with 9+8 (SCQ at center)+9 cavities, and Cryo-string-unit {long-string with 3 x 4 cavities (154 m), and short-string 3 x 3 cavities(116 m)}.

## 5.2 Civil engineering cost estimate validity

As the second point, the international review group was also concerned about the risks and cost uncertainties in civil engineering (CE). As described in Section 3.3, the ILC250 CE design was developed in collaboration with several industrial partners and the regional institutes associated with the specific site under consideration.

The costing of the ILC tunnel strictly follows the official Japanese national guidelines for estimating the construction cost of tunnels, which are generally applied to road and rail tunnels. These guidelines are updated annually by the Ministry of Land, Infrastructure and Transport [20], and present standard civil engineering design structures for various geological and rock conditions. The guidelines also provide detailed criteria for civil engineering projects, including material costs, standard speeds for the excavation methods used, time and labor costs for each process, and construction equipment costs. The CE construction plan was also reviewed by the Japan Society of Civil Engineers (JSCE), concluding the plan to be technically appropriate, as also described in Section 3.3 [21].

The total estimated cost is the bottom-up sum of these costs. It is a robust estimate as it is used as a reference price for the tendering process for civil engineering projects in Japan. Our survey of the results of CE work on public tunnels in Japan over the past 12 years shows that actual costs were within −10% and +20% of the estimated cost in 80% of cases. Cost overruns outside this range were related to unforeseen geological problems. The candidate site for the ILC has been carefully selected for its excellent geological conditions to minimize CE risks (which is not possible for transportation tunnels). Existing surveys at the ILC candidate site, although still limited in number, indicate that the chosen site is geologically favorable, as evaluated by the JSCE technical review committee described above.

## 5.3 Positron source technology development

The third point of concern for the international review group was the choice of the baseline positron source. The undulator-based polarized positron source, rather than an electron-driven source, has been the baseline since the ILC TDR, despite needing further technical development during the R&D and preparatory phases. This choice is motivated by the unique physics opportunities that simultaneous polarization of both positron and electron beams offer at the ILC. A rigorous R&D plan for the undulator-based solution and electron-driven solution are now being pursued in the context of the ILC Technology Network (ITN). The electron-driven solution is retained as a backup in case of significant delays in the engineering realization of the baseline choice. The baseline choice is to be confirmed or revised prior to the completion of the ILC engineering design.

## 6. Summary

The cost estimate for the ILC at c.m.e. of 250 GeV has been updated based on 2024 pricing. Compared to the previous evaluation conducted in 2017, components contributing to more than 70% of the total cost have been re-estimated, using new information from industrial partners and collaborating laboratories obtained in 2024. The remaining cost has been updated by scaling of the TDR cost reflecting appropriate inflation and currency fluctuation factors.

The updated 2024 cost for the 250 GeV machine is 6.78 billion ILCU for the accelerator and conventional facilities, and 196 billion JPY for the civil engineering. Compared to the 2017 estimate, the accelerator and conventional facility cost increased by approximately 60%, with around 35% attributed to worldwide inflation. The remaining increase reflects the change in production scaling, averaging multiple vendor cost-estimates, exchange rate variation, design updates, and other smaller changes. The increase of the civil engineering cost is about 50%, where around 30% is due to inflation in Japan and the remainder is due to recent design updates and enhancements.

The ILC cost uncertainty (premium) is evaluated to be approximately 30%. This value accounts for cost premiums linked to procurement challenges. The uncertainty associated with potential technical risks will decrease with further R&D and engineering studies, and it needs to be separately evaluated.

## Acknowledgements

Information and support provided by industrial partners and collaborating laboratories during this work in 2024 were essential and are deeply acknowledged for this cost estimate update. We greatly appreciate the contributions from all members of the ILC IDT Working Group 2 and the ILC Technology Network. Special advice given by L. Evans and especially important advice from S. Berry, D. Delikaris, T. Dohmae, J. Gao, C. Madec, T. Matsumoto, O. Napoly, M. Omet, J. Osborne, T. Sanuki, K-M. Schirm, R. Ueki, H. Weise and J. Zhai are particularly acknowledged. We would thank the International Cost Review Group for their constructive discussions and valuable suggestions.